\documentclass[fleqn,usenatbib]{mnras}

\usepackage{newtxtext,newtxmath}
\usepackage[T1]{fontenc}
\usepackage{ae,aecompl}

\usepackage{graphicx}
\usepackage{amsmath}
\usepackage{bm}
\usepackage{comment}
\usepackage[dvipsnames]{xcolor}
\usepackage{hyperref}

\newcommand{\Msun}{\mathrm{M}_{\sun}}
\newcommand{\Mpc}{\mathrm{Mpc}}

\newcommand{\hMpc}{h^{-1} \, \mathrm{Mpc}}

\newcommand{\Omegam}{\Omega_\mathrm{m}}

\newcommand{\TNG}{\mathrm{TNG}}
\newcommand{\TNGDark}{\mathrm{TNG}\text{-}\mathrm{Dark}}

\title[$\kappa$TNG: Baryonic Effects on WL]
{$\kappa$TNG: Effect of Baryonic Processes on Weak Lensing with IllustrisTNG Simulations}

\author[K. Osato et al.]{
Ken Osato,$^{1}$\thanks{E-mail: ken.osato@iap.fr}
Jia Liu,$^{2,3}$
and Zolt\'an Haiman$^{4}$
\\
$^{1}$Institut d'Astrophysique de Paris, Sorbonne Universit\'e, CNRS, UMR 7095, 75014 Paris, France\\
$^{2}$Berkeley Center for Cosmological Physics, University of California, Berkeley, CA 94720, USA\\
$^{3}$Lawrence Berkeley National Laboratory, 1 Cyclotron Road, Berkeley, CA 93720, USA\\
$^{4}$Department of Astronomy, Columbia University, New York, NY 10027, USA
}

\date{Accepted XXX. Received YYY; in original form ZZZ}

\pubyear{2020}

\begin{document}
\label{firstpage}
\pagerange{\pageref{firstpage}--\pageref{lastpage}}
\maketitle

\begin{abstract}
We study the effect of baryonic processes on weak lensing (WL) observables
with a suite of mock WL maps, the $\kappa$TNG,
based on the cosmological hydrodynamic simulations IllustrisTNG.
We quantify the baryonic effects on the WL angular power spectrum,
one-point probability distribution function (PDF), and number counts of peaks and minima.
We also show the redshift evolution of the effects,
which is a key to distinguish the effect of baryons
from fundamental physics such as dark energy, dark matter, and massive neutrinos.
We find that baryonic processes reduce the small-scale power,
suppress the tails of the PDF, peak and minimum counts,
and change the total number of peaks and minima.
We compare our results to existing semi-analytic models and hydrodynamic simulations,
and discuss the source of discrepancies.
The $\kappa$TNG suite includes 10,000 realisations of $5 \times 5 \, \mathrm{deg}^2$ maps
for 40 source redshifts up to $z_s = 2.6$,
well covering the range of interest for existing and
upcoming weak lensing surveys.
We also produce the $\kappa$TNG-Dark suite of maps,
generated based on the corresponding dark matter only IllustrisTNG simulations.
Our mock maps are suitable for developing analytic models that incorporate the effect of baryons,
but also particularly useful for studies that rely on mass maps,
such as non-Gaussian statistics and machine learning with convolutional neural networks.
The suite of mock maps is publicly available at Columbia Lensing (\url{http://columbialensing.org}).
\end{abstract}

\begin{keywords}
large-scale structure of Universe -- gravitational lensing: weak -- methods: numerical
\end{keywords}



\section{Introduction}
\label{sec:introduction}
Weak gravitational lensing~\citep[WL; for reviews, see][]{Bartelmann2001,HoekstraJain2008,Kilbinger2015,Mandelbaum2018b}
is a promising cosmological probe of fundamental physics
such as the nature of dark energy and dark matter, theory of gravity, and the mass sum of neutrinos.
By measuring the distortion in shapes of background galaxies caused by the foreground matter,
we can infer the intervening large-scale structure.
Ongoing Stage-III WL surveys, such as the Subaru Hyper Suprime-Cam \citep[HSC;][]{Aihara2018,Mandelbaum2018a,Hikage2019},
Dark Energy Survey \citep[DES;][]{DES2016,DES2018cosmology},
and Kilo Degree Survey \citep[KiDS;][]{deJong2015,Hildebrandt2017,Heymans2020},
have already achieved competitive cosmological constraints.
Stage-IV cosmological surveys, such as the Vera C. Rubin Observatory Legacy Survey of Space and Time \citep[LSST;][]{LSST2009},
and large-area surveys by \textit{Euclid}~\citep{Amendola2018}, and the
Nancy Grace Roman Space Telescope\footnote{\url{https://roman.gsfc.nasa.gov}},
are expected to revolutionise our understanding of the cosmos as well as fundamental physics,
enabled by their large sky coverage and deep photometry.

In order to realise the full potential of ongoing and upcoming WL surveys,
we urgently need to improve our understanding of baryonic physics~\citep[see a recent review by][]{Chisari2019}.
Baryonic processes during galaxy formation, such as radiative cooling,
feedback from black hole accretion, star formation, and supernova feedback,
redistribute the gas inside a halo and reshape its gravitational potential.
Signals from these astrophysical processes can mimic those expected from varying cosmological parameters,
causing biases in our parameter inference if left untreated
\citep{Rudd2008,Semboloni2011,Zentner2013,Mohammed2014,Osato2015,Harnois-Deraps2015,Huang2019}.
At low redshift relevant to most WL surveys,
the effects of baryons are predominately due to black hole activities,
manifested as a $\gtrsim 10\%$ level suppression in clustering around Mpc scale,
though the exact level and its redshift- and scale-dependence remains ill-constrained.
As a result, current WL surveys often need to apply aggressive scale cuts
to mitigate contamination from baryonic effects.
For example, DES applied lower limits as large as a few $\times 10 \, \mathrm{arcmin}$
in some redshift bins in analysing their Y1 data,
discarding well-measured smaller-scale data~\citep{DES2018cosmology}.
This strategy will not be sustainable for Stage-IV surveys,
which will provide high-precision measurements on sub-$\mathrm{arcmin}$ scales.
Recently, \citet{Huang2020} have extended their analysis to smaller scales with the same DES Y1 data,
by modelling and marginalising over baryonic feedback using a principal component analysis (PCA) method
based on 11 hydrodynamic simulations.
With a $2.5 \, \mathrm{arcmin}$ scale cut,
they already found a 20\% improvement on the $S_8=\sigma_8(\Omegam/0.3)^{0.5}$ constraint.

Existing works in modelling the baryonic effects mainly focus
on their impact on the two-point statistics --- the correlation function,
or its Fourier transform, the power spectrum --- of the 3D matter distribution and of the 2D WL shear or convergence.
These approaches incorporate the baryonic effects as free parameters that either alter the halo mass-concentration relation
based on the halo model~\citep{Cooray2002,Yang2013,Mead2015}, modify the gas, stellar, and dark matter density components
separately \citep{Schneider2019,Schneider2020,Arico2020a},
quantify the changes on the full power spectrum as PCAs \citep{Eifler2015,Huang2019},
or displace the matter following a pressure-like potential
\citep{Dai2018}. All these methods are calibrated against observations
and/or hydrodynamic simulations.

Recently, it has been recognised that non-Gaussian WL statistics contain rich information
beyond the traditional two-point statistics and that they will be a powerful tool
in constraining cosmological parameters.
This has been partly demonstrated on Stage-II and Stage-III surveys
\citep{Liu2015,Liux2015,Kacprzak2016,Shan2018,Martinet2018}.
However, the study of baryonic effects on non-Gaussian statistics has been limited
mainly due to the lack of analytic theories.
Therefore, these studies typically rely on simulated WL maps
to investigate baryonic effects on various non-Gaussian statistics:
three-point correlation \citep{Semboloni2013,Foreman2020,Arico2020b},
peak statistics \citep{Yang2013,Osato2015}, and minima counts \citep{Coulton2020}.

Hydrodynamic simulations are currently the state-of-the-art tool to study baryonic physics.
In these simulations, baryonic processes are modelled with subgrid prescriptions,
enabling us to capture the nonlinear astrophysics over a wide dynamic range \citep[for reviews, see][]{Somerville2015,Vogelsberger2020}.
Mock WL maps generated from hydrodynamic simulations are a critical component in studying the effect of baryons on WL observables.

In this paper, we introduce $\kappa$TNG, a suite of mock WL maps generated from the IllustrisTNG simulations.
We quantify the baryonic effects on WL statistics
including the angular power spectrum, peak counts, counts of minima, and the probability distribution function (PDF).
We compare them to those measured from the corresponding dark matter only (DMO) simulations, $\kappa$TNG-Dark,
which adopts the same initial condition but without baryonic physics.
In addition, we compare our results to existing analytic models as well as other hydrodynamic simulations.

This paper is organised as follows.
In \S~\ref{sec:simulations}, we describe our numerical methods to generate
mock WL maps from the existing IllustrisTNG simulations.
We present in \S~\ref{sec:results} our results of the baryonic effects on WL statistics
by comparing the $\kappa$TNG and $\kappa$TNG-Dark pairs.
We show their redshift dependence as well as comparisons to existing models and other hydrodynamic simulations.
We summarise our main conclusions in \S~\ref{sec:conclusions}.

\section{Numerical Simulations}
\label{sec:simulations}
In this section, we briefly describe the underlying IllustrisTNG hydrodynamic simulations
and the ray-tracing methodology we follow to generate the $\kappa$TNG mock WL maps.
Throughout this paper, we adopt a flat $\Lambda$-cold dark matter Universe
at the \textit{Planck} 2015 cosmology~\citep{Planck2015XIII}, as used in the IllustrisTNG simulations, with
Hubble constant $H_0 = 67.74 \, \mathrm{km} \, \mathrm{s}^{-1} \, \Mpc^{-1}$,
baryon density $\Omega_\mathrm{b} = 0.0486$,
matter density $\Omega_\mathrm{m} = 0.3089$,
spectral index of scalar perturbations $n_\mathrm{s} = 0.9667$,
and amplitude of matter fluctuations at $8\,\hMpc$ $\sigma_8 = 0.8159$.
We assume massless neutrinos, with an effective number of neutrino species $N_\mathrm{eff} = 3.046$.

\subsection{IllustrisTNG Hydrodynamic Simulations}

Because galaxy formation involves a wide dynamic range --- from structures internal
to stars to beyond the virial radii of the largest haloes --- it is impossible to resolve
all the relevant physical processes simultaneously in a cosmological simulation.
Therefore, baryonic processes are typically approximated by subgrid prescriptions,
in which empirical recipes are implemented to inject or remove energy and momentum
from regions of the simulation box when certain conditions are met.
Recent cosmological hydrodynamic simulations, such as the Horizon-AGN~\citep{Dubois2014},
EAGLE~\citep{Schaye2015}, BAHAMAS~\citep{McCarthy2017,McCarthy2018},
Illustris~\citep{Vogelsberger2014}, and
its successor IllustrisTNG simulations~\citep{Nelson2019,Pillepich2018,Nelson2018,Springel2018,Naiman2018,Marinacci2018}
have already provided invaluable insights into the impact of baryons on cosmological observables.

In this work, we use the IllustrisTNG simulations as our base.
They are a set of cosmological, large-scale gravity and magneto-hydrodynamical simulations
generated with the moving mesh code AREPO~\citep{Springel2010}.
Subgrid prescriptions are incorporated to model stellar evolution, chemical enrichment, gas cooling,
and black hole feedback~\citep{Pillepich2018}.
The full simulation set includes three box sizes, each with simulations of different resolutions.
Here, we employ the highest-resolution simulation for the largest box, TNG300-1 (hereafter TNG),
which covers a comoving volume of $(205\,\hMpc)^3$.
It has a mass resolution of $7.44 \times 10^6 \, h^{-1}\Msun$ and $3.98 \times 10^7 \, h^{-1}\Msun$
for initial gas and dark matter particles, respectively.
To discriminate baryonic effects,
we make use of the corresponding DMO simulations TNG300-1-Dark (hereafter TNG-Dark),
which has a mass resolution $4.73 \times 10^7 \, h^{-1} \, \Msun$ for DM particles.
TNG and TNG-Dark have the same initial conditions and only differ in the inclusion of baryonic physics.
These simulations accurately resolve the small-scale, nonlinear gravitational evolution
of the matter density field that is accessible by Stage-IV cosmological surveys.

To study the WL observables, we need to generate mock maps with a reasonable area (at least a few $\mathrm{deg}^2$)
and redshift coverage (up to $z\approx$ 2).
In addition, a large number of realisations is important to suppress cosmic variance.
The TNG simulation snapshots are outputted at 99 redshifts between $z=0\text{--}20$,
allowing one to build light cones that are necessary to construct mock WL maps.
However, since IllustrisTNG was not tailored to build mock WL maps,
we must overcome some difficulties to achieve our goal.
First, these snapshots are not output at regular intervals of comoving distance,
as would have been required for standard WL ray-tracing schemes~\citep{dietrich2010,Liu2018MassiveNuS,SLICS,cosmoSLICS}.
Second, the box size, despite being very large for hydrodynamic simulations,
is relatively small compared to the usual DMO cosmological simulations;
and hence it can only cover a small patch of the sky at high redshifts.
Finally, due to the high computational cost, only one simulation was generated at the box size and resolution we need.
We adapt our ray-tracing method to accommodate these pre-existing settings, which we describe in detail next.

\subsection{$\kappa$TNG: Ray-Traced Weak Lensing Mock Maps}

\begin{figure*}
\includegraphics[width=\textwidth]{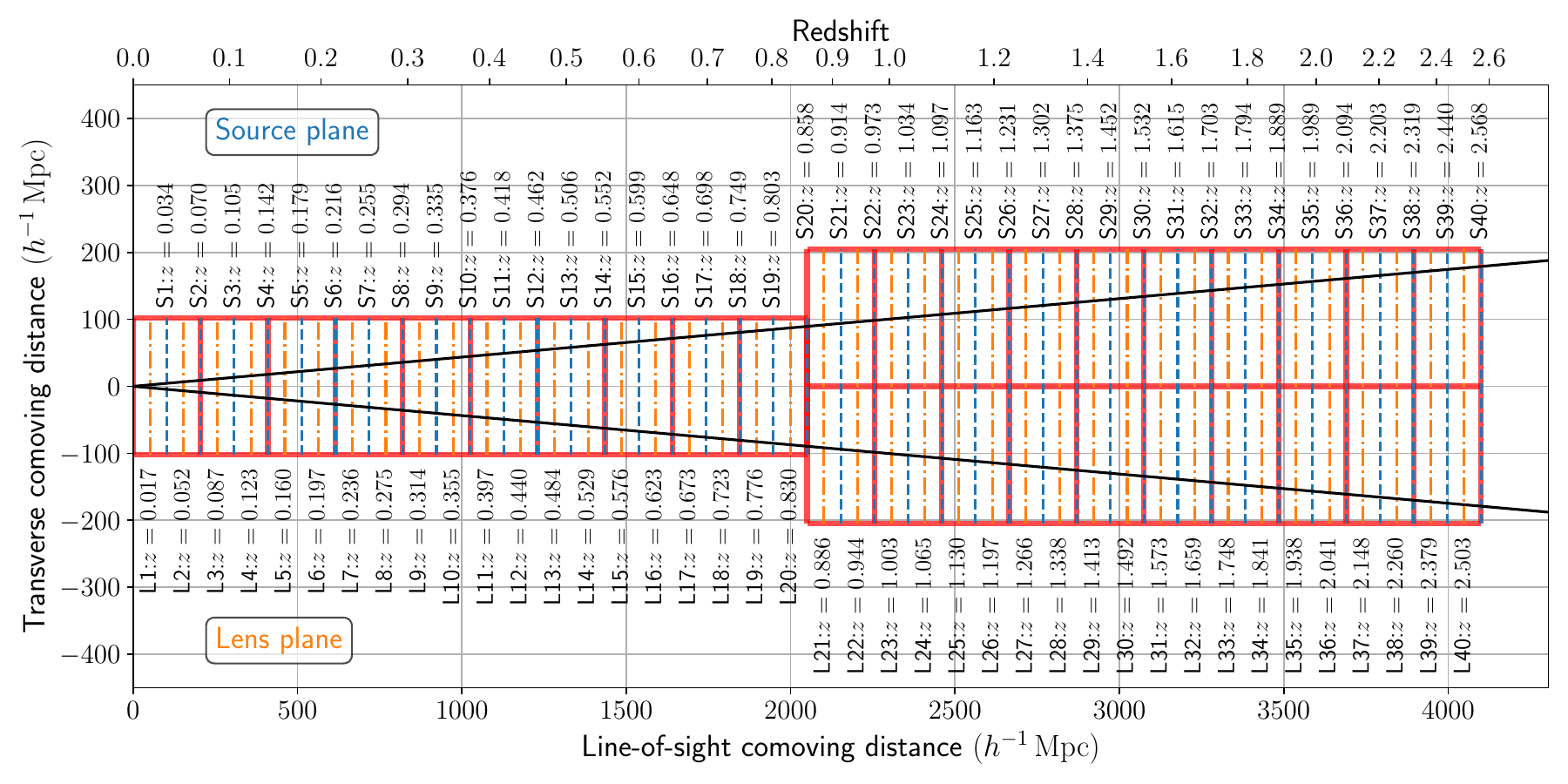}
\caption{The construction of our light cone.
Each red box corresponds to one snapshot.
The last 10 snapshots are replicated 4 times (twice in each transverse direction) to increase the solid area to cover $5\times5$ degrees.
Each box is divided into 2 lens planes of $\Delta \chi = 102.5 \, \hMpc$ thickness.
We list the redshifts at the central positions of source (blue dashed lines) and lens (orange dash-dotted lines) planes.
The black lines correspond to our opening angle of $5 \, \mathrm{deg}$. We generate lensing mocks at 40 source redshifts.}
\label{fig:light_cone}
\end{figure*}

To build a light cone, we stack the TNG snapshots
along the line-of-sight in a fixed interval of $205 \, \hMpc$ ($=$ TNG box size).
The comoving distance between $z = 0$ and $z \approx 2.5$ is approximately $4000 \, \hMpc$,
which requires 20 boxes in total to cover.
We select the TNG snapshots that are the closest to the centres of the 20 light cone intervals,
if they were equally spaced
in comoving distance.\footnote{In principle, we could use all 100 TNG snapshots of in generating our light cone.
However, we expect negligible differences between 2--3 consecutive snapshots
and hence we limited the number of snapshots to 20 in total for faster data transfer and computation.}
We summarise our light cone configuration in Table~\ref{tab:snapshots},
including the comoving distance $\chi$ and redshift for each snapshot, as well as the corresponding TNG file.
We illustrate our light cone configuration in Figure~\ref{fig:light_cone}.
The opening angle of the light cone is $5 \times 5 \, \mathrm{deg}^2$.
The last 10 snapshots are replicated 4 times in the transverse direction,
so that we can cover the desired map size ($5\times5 \, \mathrm{deg}^2$) at the highest-redshift maps.
Our light cone extends up to redshift $z = 2.57$. We construct mock WL maps at 40 source redshifts using this light cone.

\begin{table}
  \caption{Summary of our light cone. Columns: (1) snapshot number;
  (2) comoving distance and (3) redshift at the snapshot centre;
  (4) corresponding TNG snapshot redshift and (5) TNG file number.}
  \label{tab:snapshots}
\begin{center}
\begin{tabular}{ccccc}
Snapshot & $\chi$ ($\hMpc$) & $z$ &
$z^{\TNG}$  & TNG File \\
\hline \hline
1 & $102.5$ & $0.034$ & $0.03$ & 96 \\
2 & $307.5$ & $0.105$ & $0.11$ & 90 \\
3 & $512.5$ & $0.179$ & $0.18$ & 85 \\
4 & $717.5$ & $0.255$ & $0.26$ & 80 \\
5 & $922.5$ & $0.335$ & $0.33$ & 76 \\
6 & $1127.5$ & $0.418$ & $0.42$ & 71 \\
7 & $1332.5$ & $0.506$ & $0.50$ & 67 \\
8 & $1537.5$ & $0.599$ & $0.60$ & 63 \\
9 & $1742.5$ & $0.698$ & $0.70$ & 59 \\
10 & $1947.5$ & $0.803$ & $0.79$ & 56 \\
11 & $2152.5$ & $0.914$ & $0.92$ & 52 \\
12 & $2357.5$ & $1.034$ & $1.04$ & 49 \\
13 & $2562.5$ & $1.163$ & $1.15$ & 46 \\
14 & $2767.5$ & $1.302$ & $1.30$ & 43 \\
15 & $2972.5$ & $1.452$ & $1.41$ & 41 \\
16 & $3177.5$ & $1.615$ & $1.60$ & 38 \\
17 & $3382.5$ & $1.794$ & $1.82$ & 35 \\
18 & $3587.5$ & $1.989$ & $2.00$ & 33 \\
19 & $3792.5$ & $2.203$ & $2.21$ & 31 \\
20 & $3997.5$ & $2.440$ & $2.44$ & 29 \\
\hline
\end{tabular}
\end{center}
\end{table}

To model the propagation of light rays in our light cone, we employ a multiple lens plane approximation
\citep{Blandford1986,Seitz1994,Jain2000,Vale2003,Hilbert2009},
where the smooth matter distribution is approximated as discrete density planes of thickness $\Delta \chi$.
Light rays originating from the observer at $z=0$ are deflected only at the planes and travel in straight lines between the planes.
At the angular position $\bm{\beta}^k$ at the $k^\mathrm{th}$ plane,
the deflection angle $\bm{\alpha}$ is the gradient of the 2D lensing potential $\psi$:
\begin{align}
  \bm{\alpha}^k \left( \bm{\beta}^k \right) =
  \nabla_{\bm{\beta}^k} \psi^k \left( \bm{\beta}^k \right) .
\end{align}
The lensing potential can be computed from the Poisson equation,
\begin{align}
  \label{eq:Poisson}
  \nabla_{\bm{\beta}^k}^2 \psi^k \left( \bm{\beta}^k \right) =
  2 \sigma^k \left( \bm{\beta}^k \right),
  \end{align}
 where the dimensionless surface density $\sigma^k$ is the projected matter distribution of the $k^\mathrm{th}$ lens plane,
  \begin{align}
      \sigma^k \left( \bm{\beta}^k \right) =& \frac{3 H_0^2 \Omegam}{2 c^2}
  \frac{\chi^k}{a^k}
 \int_{\chi^k-\Delta \chi/2}^{\chi^k+\Delta \chi/2}
  \delta \left( \bm{\beta}^k, \chi' \right) d\chi,
\end{align}
where $c$ is the speed of light, $a$ is the scale factor,
and $\delta=(\rho-\bar\rho)/\bar\rho$ is the three-dimensional matter overdensity.
The lensed position $\bm{\beta}^k$ is
the sum of previous deflection angles,
\begin{align}
\bm{\beta}^k (\bm{\theta}) = \bm{\theta} - \sum_{i=1}^{k-1}
\frac{\chi^k-\chi^i}{\chi^k} \bm{\alpha}^i \left( \bm{\beta}^i \right),
\ (k = 2, 3, \ldots)
\end{align}
where we impose $\bm{\beta}^{0} = \bm{\beta}^{1} = \bm{\theta}$ for initial rays.
By differentiating $\bm{\beta}^k$ with respect to $\bm{\theta}$,
we can obtain the distortion matrix that maps the initial position to the lensed position $\bm{\theta} \to \bm{\beta}$,
\begin{align}
A_{ij} (\bm\theta, \chi) \equiv & \frac {\partial \beta_i (\bm\theta, \chi)} {\partial \theta_j}
\\ \nonumber
= &
\begin{pmatrix}
1 - \kappa - \gamma_1 & -\gamma_2 + \omega \\
-\gamma_2 - \omega & 1 - \kappa + \gamma_1
\end{pmatrix},
\end{align}
where, for legibility, we suppressed the dependence on $(\bm\theta, \chi)$ for
the convergence $\kappa$, the shear components $\gamma_1$ and $\gamma_2$, and the rotation $\omega$.
To obtain $\bm{\beta}$ and $A_{ij}$,
we employ a memory-efficient ray-tracing scheme developed in \citet{Hilbert2009},
where the computation for the $k^\mathrm{th}$ plane
depends only on the quantities at $(k-1)^\mathrm{th}$ and $(k-2)^\mathrm{th}$ planes.

The configuration of our source planes and lens planes are illustrated in Figure~\ref{fig:light_cone}.
We also summarise the corresponding comoving distances and redshifts in Table~\ref{tab:configuration}.
We divide each snapshot into 2 density slices of thickness $\Delta \chi = 102.5 \, \hMpc$.
We then project each slice onto a regular grid of $4096^2$ pixels.
The dark matter, gas, and star particles (in the case of TNG-Dark, only dark matter particles) are assigned
to the grid with a triangular-shaped cloud scheme.
Next, we apply a fast Fourier transform (FFT) to Eq.~\eqref{eq:Poisson}
to obtain the derivatives of the lensing potential.
We evaluate the deflection angle and
the distortion matrix at the angular position $\bm{\beta}^k$
and generate ($\kappa$, $\gamma_1$, $\gamma_2$, $\omega$) maps at each source plane.
Each map is on a regular grid with $1024^2$ pixels,
corresponding to a pixel size of $0.29 \, \mathrm{arcmin}$.
In total, we generate 40 source planes.
Note that $i^\mathrm{th}$ lens plane contains the contribution from the range
$(\chi^{i}-\Delta \chi/2, \chi^{i}+\Delta \chi/2)$,
but observables (e.g., $\bm{\beta}^{i}$) are defined at $\chi^{i}$.
Thus, the $i$-th source plane, where the observables are output,
is placed at $\chi^{i}+\Delta \chi/2$.

\begin{table}
  \caption{The comoving distance $\chi$ and redshifts ($z_s$, $z_l$) for our source and lens planes.
  The light cone configuration is illustrated in Figure~\ref{fig:light_cone}.}
  \label{tab:configuration}
\begin{tabular}{ccc|ccc}
Source & $\chi$ & &
Lens & $\chi$ &  \\
Plane & ($\hMpc$) & $z_s$ &
Plane & ($\hMpc$) & $z_l$ \\
\hline \hline
S1 & $102.5$ & $0.034$ & L1 & $51.25$ & $0.017$ \\
S2 & $205.0$ & $0.070$ & L2 & $153.75$ & $0.052$ \\
S3 & $307.5$ & $0.105$ & L3 & $256.25$ & $0.087$ \\
S4 & $410.0$ & $0.142$ & L4 & $358.75$ & $0.123$ \\
S5 & $512.5$ & $0.179$ & L5 & $461.25$ & $0.160$ \\
S6 & $615.0$ & $0.216$ & L6 & $563.75$ & $0.197$ \\
S7 & $717.5$ & $0.255$ & L7 & $666.25$ & $0.236$ \\
S8 & $820.0$ & $0.294$ & L8 & $768.75$ & $0.275$ \\
S9 & $922.5$ & $0.335$ & L9 & $871.25$ & $0.314$ \\
S10 & $1025.0$ & $0.376$ & L10 & $973.75$ & $0.355$ \\
S11 & $1127.5$ & $0.418$ & L11 & $1076.25$ & $0.397$ \\
S12 & $1230.0$ & $0.462$ & L12 & $1178.75$ & $0.440$ \\
S13 & $1332.5$ & $0.506$ & L13 & $1281.25$ & $0.484$ \\
S14 & $1435.0$ & $0.552$ & L14 & $1383.75$ & $0.529$ \\
S15 & $1537.5$ & $0.599$ & L15 & $1486.25$ & $0.576$ \\
S16 & $1640.0$ & $0.648$ & L16 & $1588.75$ & $0.623$ \\
S17 & $1742.5$ & $0.698$ & L17 & $1691.25$ & $0.673$ \\
S18 & $1845.0$ & $0.749$ & L18 & $1793.75$ & $0.723$ \\
S19 & $1947.5$ & $0.803$ & L19 & $1896.25$ & $0.776$ \\
S20 & $2050.0$ & $0.858$ & L20 & $1998.75$ & $0.830$ \\
S21 & $2152.5$ & $0.914$ & L21 & $2101.25$ & $0.886$ \\
S22 & $2255.0$ & $0.973$ & L22 & $2203.75$ & $0.944$ \\
S23 & $2357.5$ & $1.034$ & L23 & $2306.25$ & $1.003$ \\
S24 & $2460.0$ & $1.097$ & L24 & $2408.75$ & $1.065$ \\
S25 & $2562.5$ & $1.163$ & L25 & $2511.25$ & $1.130$ \\
S26 & $2665.0$ & $1.231$ & L26 & $2613.75$ & $1.197$ \\
S27 & $2767.5$ & $1.302$ & L27 & $2716.25$ & $1.266$ \\
S28 & $2870.0$ & $1.375$ & L28 & $2818.75$ & $1.338$ \\
S29 & $2972.5$ & $1.452$ & L29 & $2921.25$ & $1.413$ \\
S30 & $3075.0$ & $1.532$ & L30 & $3023.75$ & $1.492$ \\
S31 & $3177.5$ & $1.615$ & L31 & $3126.25$ & $1.573$ \\
S32 & $3280.0$ & $1.703$ & L32 & $3228.75$ & $1.659$ \\
S33 & $3382.5$ & $1.794$ & L33 & $3331.25$ & $1.748$ \\
S34 & $3485.0$ & $1.889$ & L34 & $3433.75$ & $1.841$ \\
S35 & $3587.5$ & $1.989$ & L35 & $3536.25$ & $1.938$ \\
S36 & $3690.0$ & $2.094$ & L36 & $3638.75$ & $2.041$ \\
S37 & $3792.5$ & $2.203$ & L37 & $3741.25$ & $2.148$ \\
S38 & $3895.0$ & $2.319$ & L38 & $3843.75$ & $2.260$ \\
S39 & $3997.5$ & $2.440$ & L39 & $3946.25$ & $2.379$ \\
S40 & $4100.0$ & $2.568$ & L40 & $4048.75$ & $2.503$ \\
\hline
\end{tabular}
\end{table}

To suppress sample variance, we need a large number of map realisations.
To accomplish this goal with only one simulation of TNG, we exploit the periodic boundary condition
and the fact that the light cone does not cover the whole simulation box.
We construct \textit{pseudo-independent} map realisations by randomly translating and rotating the snapshots.
First, for each snapshot,
we (1) translate all particles by a random number along each of the 3 axes,
(2) rotate the snapshot by $0$, $90$, $180$, or $270$ degrees around each of the three axes,
and (3) apply a random flip along any of the 3 axes.
We repeat this process 100 times and generate lensing potential planes from these randomised snapshots.
Next, for each lens plane set, we further translate and rotate
the planes randomly, for 100 times.
Finally, we obtain 10,000 pseudo-independent realisations.
Such procedure has been studied thoroughly in the past by \citet{Petri2016},
who found that even with only one simulation,
the snapshots can be repeatedly recycled to produce up to a few $\times 10^4$ WL map realisations
whose power spectra and high-significance peak counts can be treated as statistically independent.
Such a large number of realisations would also allow the application of machine learning
to study the mapping from the DMO to hydrodynamic WL maps.

\section{Results}
\label{sec:results}
\begin{figure*}
\includegraphics[width=0.85\textwidth]{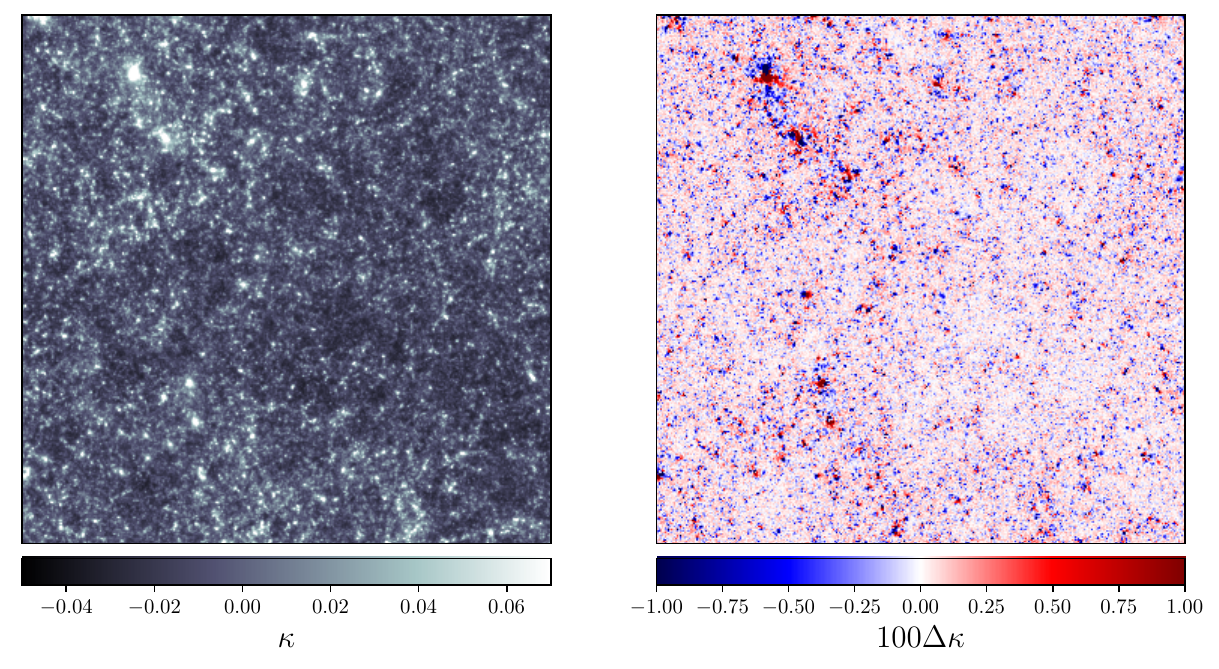}
\caption{\textbf{Left panel}: An example of $\kappa$TNG convergence map at source redshift $z_s = 1.034$.
\textbf{Right panel}: The difference between the paired $\kappa$TNG and $\kappa$TNG-Dark maps
$\Delta \kappa = \kappa^{\TNG} - \kappa^{\TNGDark}$.
They have the same initial conditions, but the baryonic processes are only included in the former.
The $\kappa$TNG-Dark map is not shown here as it is almost indistinguishable from the $\kappa$TNG map.
All our maps are $5\times5 \, \mathrm{deg}^2$ in size, each on a $1024^2$ regular grid
 corresponding to a pixel size of $0.29 \, \mathrm{arcmin}$.}
\label{fig:kappa_map}
\end{figure*}

In Figure~\ref{fig:kappa_map},
we show an example of $\kappa$TNG map at $z_s = 1.034$ (S23), as well as the difference
between the pair of hydrodynamic and DMO maps.
The amplitude of the difference appears to be correlated with the $\kappa$ values of the pixels,
but in a somewhat complex fashion.
For example, some dipole patterns are clearly seen in the overdense regions.
In most baryonification models, e.g., HMcode \citep{Mead2015}
and baryonic correction models \citep[e.g.][]{Schneider2015, Schneider2019},
the halo positions, or clustering properties, are assumed to be unchanged due to baryonic effects.
It may be caused by collimated active galactic nuclei (AGN) jets or changes in halo positions and ellipticities.
We leave a thorough investigation to future work.
Next, we quantify these effects with the power spectrum, PDF, and peak and minimum counts.
Together, they capture both Gaussian and non-Gaussian information and are complementary probes for cosmology.

\subsection{Power Spectrum}
We measure the angular power spectra of the convergence map $C_\kappa$ using a binned estimator,
\begin{align}
  C_\kappa (\ell_i) &= \frac{1}{N_i}
  \sum_{|\bm{\ell}| \in [\ell_i^\mathrm{min}, \ell_i^\mathrm{max}]}
  |\tilde{\kappa}(\bm{\ell})|^2 ,
\end{align}
where $\ell_i$ is the mean of multipoles in the $i^\mathrm{th}$ bin with bounds $[\ell_i^\mathrm{min}, \ell_i^\mathrm{max}]$,
$N_i$ is the number of modes, and $\tilde{\kappa}(\bm{\ell})$ is the Fourier transform of the convergence field,
computed on a $1024^2$ regular grid with FFT.
We assign 10 log-equally spaced bins in the range of $[10^2, 10^3]$ and
20 log-equally spaced bins in the range of $[10^3, 10^4]$ --- in total 30 multipole bins.

\begin{figure}
\includegraphics[width=\columnwidth]{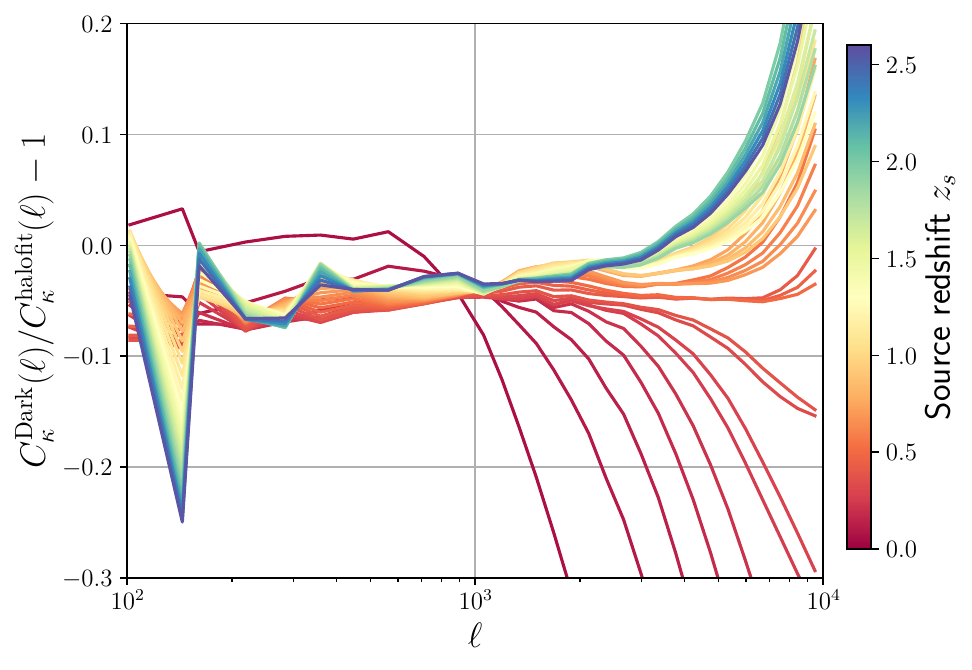}
\caption{The fractional differences of the $\kappa$TNG-Dark power spectra
with respect to the analytic \textit{halofit} predictions.
The result for the first source redshift $z_s = 0.034$ is not shown
due to large statistical uncertainty.}
\label{fig:halofit}
\end{figure}

We first validate our ray-tracing code by comparing the $\kappa$TNG-Dark power spectra with the theoretical prediction
from \textit{halofit} \citep{Smith2003,Takahashi2012}.
We show the comparison for all 40 source redshifts in Figure~\ref{fig:halofit}.
On large scales ($\ell \approx$ a few $\times 100$), our measured power spectra are lower than the theoretical prediction by $\approx 5\%$.
The reduced power reflects the missing large-scale modes due to the limited TNG box size.
However, since both the $\kappa$TNG and $\kappa$TNG-Dark maps are impacted by this effect on the same footing,
it is expected to be canceled out when we consider the differences between the simulation pair.
Note that, for lower source redshifts, the statistical variance is so large that
the measured power spectrum can be accidentally closer to the \textit{halofit} results.
On small scales ($\ell > 1000$), the differences for $z_s > 0.4$ curves are
within the known \textit{halofit} modelling uncertainty of $\approx 10 \%$.
For low redshifts $z_s < 0.4$, the discrepancies become larger,
because the same $\ell$ value corresponds to smaller structures,
some of which are below the simulation resolution.
To investigate the effect of map resolution,
we generate 50 additional high-resolution test maps with number of grids doubled at a side.
The fiducial power spectra are consistent
with these high-resolution maps within $1 \%$ for $100 < \ell < 5000$,
but are suppressed by $\approx 5 \%$ at $\ell = 10000$ for $z_s = 1.034$ (S23), and
the suppression at $\ell = 10000$ becomes larger for lower source redshift.
However, $\kappa$TNG and $\kappa$TNG-Dark maps are affected by the resolution
in the same manner, and thus, the angular power spectrum ratio is not subject to
the resolution effect.\footnote{The suppression in the power spectrum due to resolution
can be corrected by introducing a damping factor \citep{Takahashi2017}.
However, we do not apply this damping factor because it does not take into account
the redshift dependence and we are interested mainly in the angular power spectrum ratio.}
Finally, we note that the \textit{halofit} model was calibrated
against simulations with lower resolution than the TNG and
is known to over-predict the WL power on small scales,
so it should only be considered as a rough estimate in this regime
\citep[see, e.g.,][]{Mead2015,Wei2018}.

\begin{figure}
\includegraphics[width=\columnwidth]{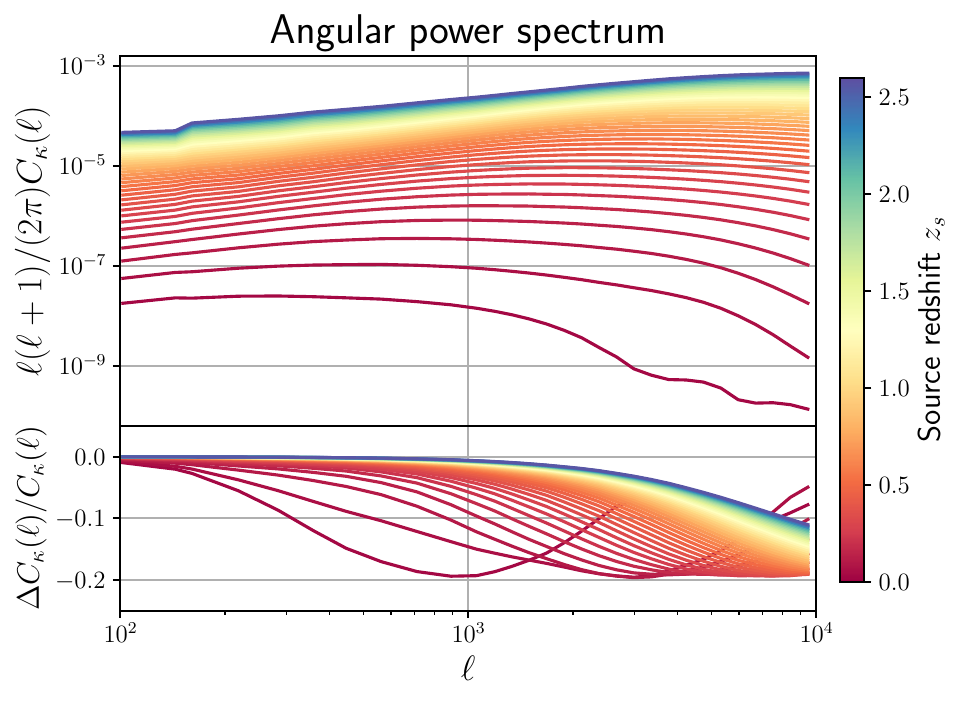}
\caption{\textbf{Upper panel}: Angular power spectra for all source redshifts from our $\kappa$TNG maps
(red to blue: low to high redshifts).
We do not show the curves for $\kappa$TNG-Dark maps,
as they are almost indistinguishable by eye from the $\kappa$TNG curves.
\textbf{Lower panel}: Fractional differences between the $\kappa$TNG and $\kappa$TNG-Dark power spectra
for all 40 source redshifts,
where $\Delta C_\kappa / C_\kappa = C_\kappa^{\kappa \TNG} / C_\kappa^{\kappa \TNGDark} - 1$.
The power spectra and ratios are means over 10,000 map realisations.
The spectra between $\ell = 3771.0 \text{--} 9465.6$
for the first source redshift $z_s = 0.034$ are not shown
because of strong artefacts due to resolution.}
\label{fig:power_all}
\end{figure}

We investigate the effect of baryons on the WL power spectrum next.
In Figure~\ref{fig:power_all}, we present the redshift evolution of the $\kappa$TNG power spectrum in the upper panel,
as well as the fractional differences between the $\kappa$TNG and $\kappa$TNG-Dark maps in the lower panel.
Both the power spectra and ratios are averaged over $N_r = 10000$
realisations\footnote{We take the mean of ratio, instead of the ratio of the mean power spectra,
to reduce the cosmic variance}.
In general, a spoon-like feature is seen.
At large scales, the ratio is close to unity, i.e., no significant baryonic effects.
On small scales, a suppression up to $20\%$ is seen.
This is mainly because feedback processes such as feedback from black hole accretion and supernova explosions,
which can remove gas from the halo centre, reduce the matter clustering on relevant scales.
The location of the dip moves from small angular scales at high redshift to larger angular scales ($\ell \approx 1000$) at lower redshift,
mainly because the same physical scales extend larger angles at lower redshift. In addition, at low redshift,
baryonic feedback is also more powerful, reaching further in physical scales.
Finally, the upturn seen at much smaller scales is the result of radiative cooling,
which enhances the matter clustering. In Appendix~\ref{sec:fitting},
we present a fitting formula for the suppression due to baryons on the angular power spectrum.

The characteristic shape of the baryonic effects on the WL power spectrum mimics the effect of changing cosmological parameters,
in particular the mass sum of neutrinos, which also shows a spoon-like feature \citep{Lesgourgues2006,Hannestad2020}.
The key to distinguish them is their different redshift- and scale-dependence.
In addition, they may also manifest differently in other statistics beyond the power spectrum, such as the ones we study next.

\subsection{Probability Distribution Function (PDF)}
Here, we present the effect of baryons on the WL one-point PDF.
Past studies of the WL PDF have shown that it has the potential to significantly tighten the cosmological constraints,
compared to using the power spectrum alone \citep{Wang2009,Liu2016pdf,Patton2017,Liu2019}.
Recently, \citet{Thiele2020} developed an analytic model for the WL one-point PDF and its auto-covariance,
based on a halo-model formalism.
While baryonic effects have not been included in these studies, future extensions could incorporate baryonic effects.

\begin{figure}
\includegraphics[width=\columnwidth]{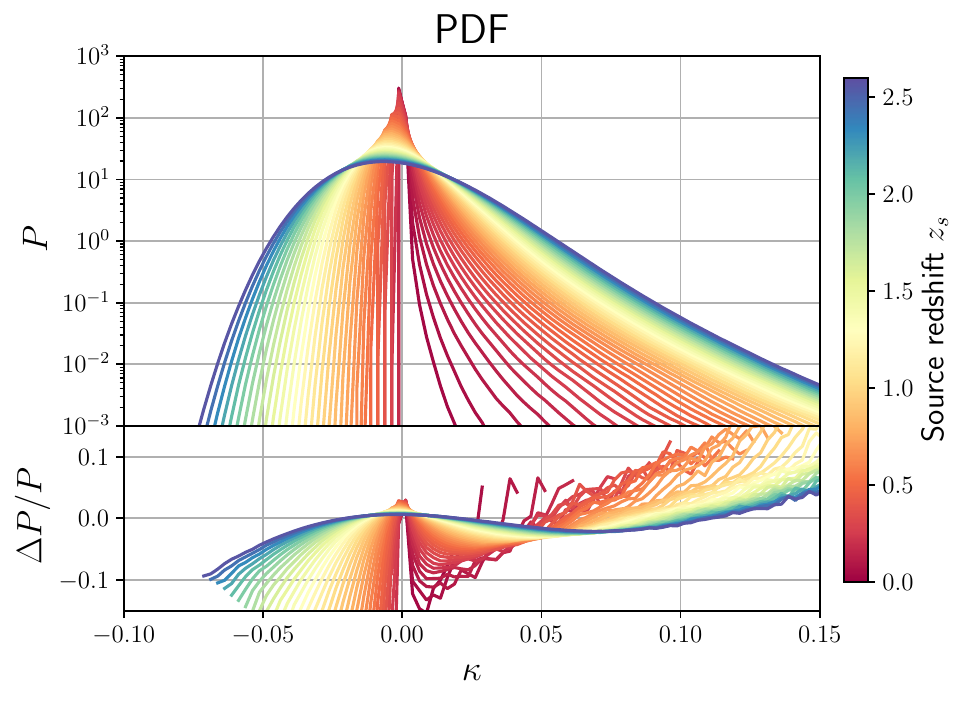}
\caption{\textbf{Upper panel}: The PDFs for all source redshifts
from our $\kappa$TNG maps (red to blue: low to high redshifts).
\textbf{Lower panel}: Fractional differences between the $\kappa$TNG and $\kappa$TNG-Dark maps for all 40 source redshifts,
where $\Delta P / P = P^{\kappa \TNG} / P^{\kappa \TNGDark} - 1$.
The maps are smoothed with a $2\, \mathrm{arcmin}$ Gaussian filter.
The PDFs and ratios are means over 10,000 map realisations.}
\label{fig:PDF}
\end{figure}

We first smooth the maps with a $\theta_\mathrm{G} = 2 \, \mathrm{arcmin}$ Gaussian window to suppress small scale noise.
The window function is given as
\begin{equation}
W (\theta) = \frac{1}{\pi \theta_\mathrm{G}^2} \exp \left( - \frac{\theta^2}{\theta_\mathrm{G}^2} \right) .
\end{equation}
We exclude pixels within $2 \theta_\mathrm{G}$ from the edge due to incomplete smoothing.
To measure the PDF, we then measure the histogram of the pixels,
binned by their $\kappa$ value with
the bin width $\Delta \kappa = 0.025$\footnote{To account for the slight (de)magnification of the ray bundles during ray-tracing,
we weigh the pixels by the inverse magnification in computing the PDF.
Since smoothing smears the small-scale structures, the PDFs from high-resolution and fiducial maps are consistent within $1 \%$.
We expect this to have negligible effect on our results.
The smoothing further reduces its effect. See more discussion in \citet{Takahashi2011}.}.
In Figure~\ref{fig:PDF}, we present the redshift evolution of the $\kappa$TNG PDF in the upper panel,
as well as the fractional differences between the $\kappa$TNG and $\kappa$TNG-Dark maps in the lower panel.
The PDF is skewed with a high $\kappa$ tail for all redshifts considered in our work,
indicating the non-Gaussian information that is not captured by the power spectrum.

The primary effect of baryonic processes is the suppression of
the intermediate positive $\kappa$ and negative $\kappa$ regions, as well as an enhancement at the very high $\kappa$ tail.
In the intermediate positive convergence regime, feedback processes tend to expel the gas from overdense regions,
resulting in a suppression of the most clustered, nonlinear structures.
Consequently, even the void regions can acquire some mass,
resulting in the observed decrease in number of highly-negative pixels.
The redistribution of matter by baryonic effects smoothes the overall density fluctuation,
resulting in more pixels with small convergence ($\kappa \approx 0$) signals.
The enhancement at the very high $\kappa$ tail is due to radiative cooling.

\subsection{Peaks and Minima}

\begin{figure*}
\includegraphics[width=\columnwidth]{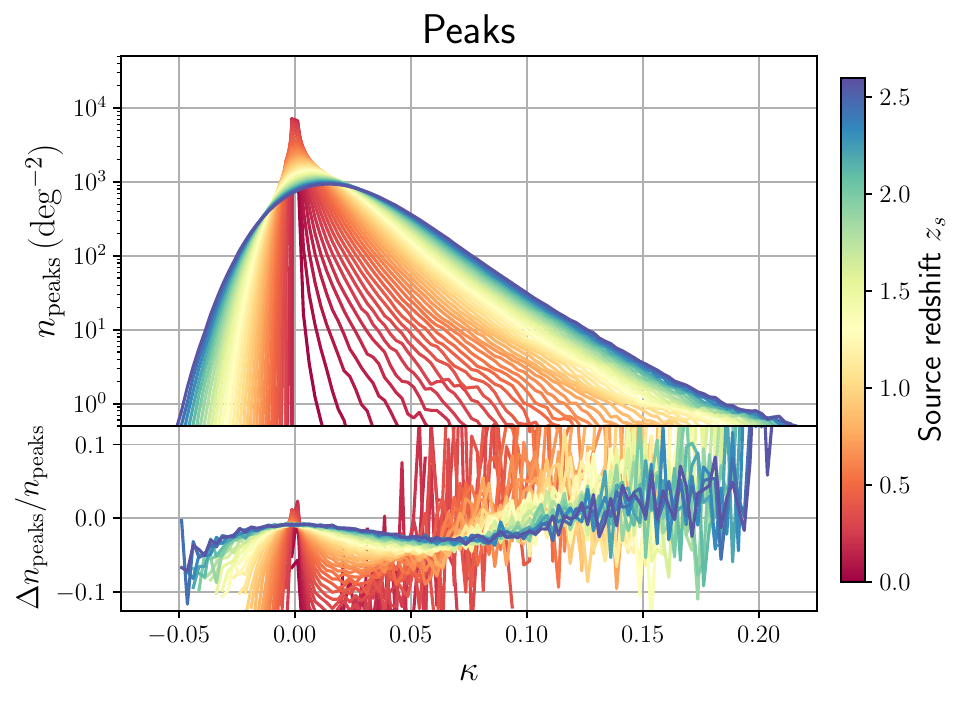}
\includegraphics[width=\columnwidth]{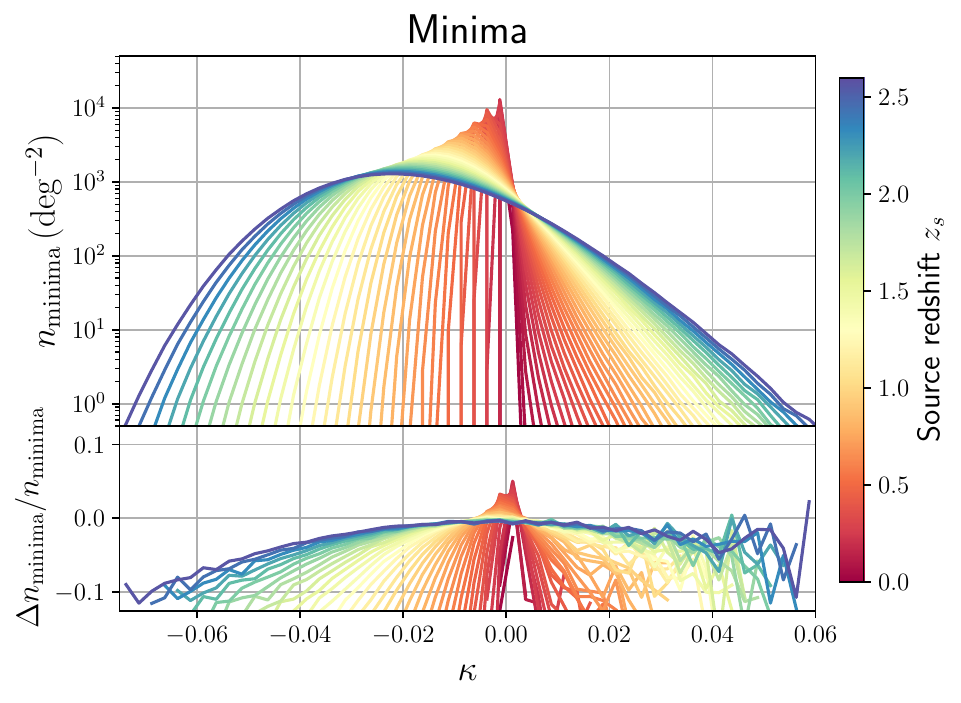}
\caption{\textbf{Left panel}: Peak counts from our $\kappa$TNG maps (upper panel) and the fractional differences
between the $\kappa$TNG and $\kappa$TNG-Dark maps (lower panel) for all 40 source redshifts
(red to blue: low to high redshifts), where $\Delta n / n = n^{\kappa \TNG} / n^{\kappa \TNGDark} - 1$.
\textbf{Right panel}: the same as the left, but for minimum counts.
The maps are smoothed with a $2\,\mathrm{arcmin}$ Gaussian filter
and the measurements are means over 10,000 map realisations.}
\label{fig:peaksminima}
\end{figure*}

The number count of WL peaks --- pixels with a higher value than their surrounding 8 pixels --- is the most well-studied
non-Gaussian statistic~\citep[e.g.][]{Kratochvil2010,dietrich2010}.
It has been applied to observational data from the CFHTLenS~\citep{Liu2015}, CS82~\citep{Liux2015}, DES~\citep{Kacprzak2016},
and KiDS surveys~\citep{Shan2018,Martinet2018} and returned comparable or better constraints to those from the two-point statistics.
The convergence peaks originate from single massive haloes or
superpositions of multiple intermediate mass haloes \citep{Yang2011,LiuHaiman2016}.
They are sensitive to both the growth of structure and the expansion history of the universe.
In contrast, WL minima --- pixels with a lower value than their surrounding 8 pixels --- also contain information
complementary to peaks \citep{Maturi2010,Coulton2020}.
WL peaks and minima are easily measurable in WL mass maps,
without the need to identify physical structures such as clusters or voids.
Past works studying the baryonic effects on WL peaks and minima have found that
they are affected differently than the power spectrum \citep{Yang2013,Osato2015,Weiss2019,Coulton2020},
in terms of biases in cosmological parameters.
Therefore, they can also serve as a tool to calibrate potential systematics.
Compared to the simulations used in previous studies,
the $\kappa$TNG mock WL maps cover a wide range of redshifts and
take advantage of the state-of-the art subgrid models developed by the TNG team.

In Figure~\ref{fig:peaksminima}, we show the number counts of peaks and minima in the upper panels,
as well as the fractional differences between the $\kappa$TNG and $\kappa$TNG-Dark maps in the lower panels.
To measure the number counts of peaks and minima, we again smooth the convergence maps
with a $\theta_\mathrm{G}=2 \, \mathrm{arcmin}$ Gaussian window.
The peaks and minima are counted in equally spaced bins with width $\Delta \kappa = 0.025$.
Similarly to PDFs, pixels within $2 \theta_\mathrm{G}$ from the edge are discarded,
and the results with high-resolution and fiducial maps are consistent within $1 \%$.
The overall baryonic effects in peaks and minimum counts are somewhat similar to that in PDF:
both the positive and negative tails are suppressed.
However, we also see an additional effect of reduced number of peaks or minima
in contrast to the PDF, which is always normalised to unity.
These findings are consistent with previous studies \citep{Osato2015,Coulton2020},
though the amplitude of the change depends on the strength of the baryonic feedback implemented in the underlying hydrodynamic simulations.
We compare our results to previous works, including both (semi-)analytic models and other hydrodynamic simulations, in the next subsection.

\subsection{Comparison with Previous Studies}

\begin{figure*}
\includegraphics[width=0.9\textwidth]{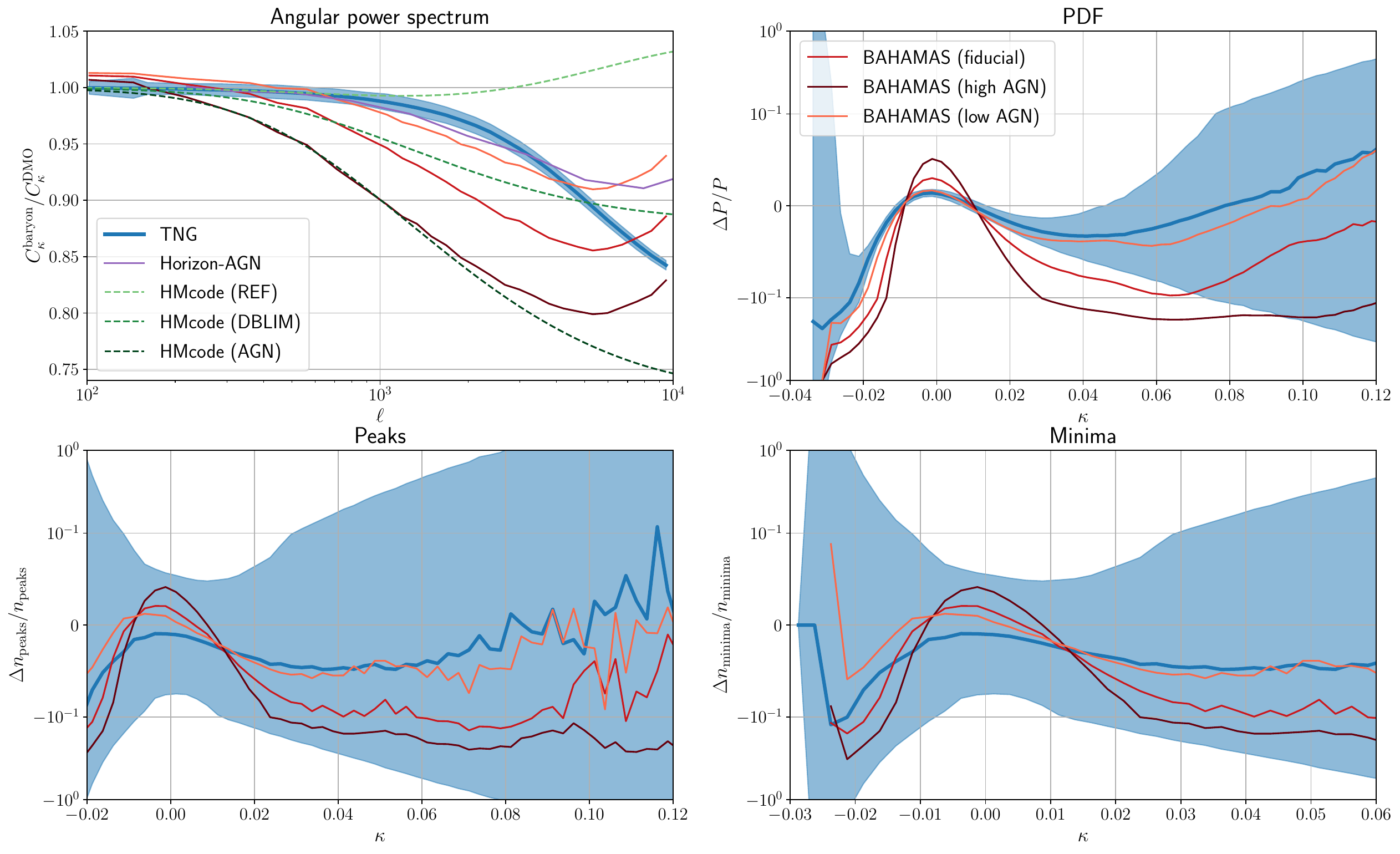}
\caption{Comparison of the effects of baryonic processes on
our angular power spectrum (\textbf{upper-left}),
one-point PDF (\textbf{upper-right}), number of peaks (\textbf{lower-left}),
and number of minima (\textbf{lower-right}) to previous works,
including analytic models (dashed curves) for the power spectrum from the HMcode and
hydrodynamic simulations (solid curves) BAHAMAS and Horizon-AGN.
All curves are shown for $z_s = 1$ except $\kappa$TNG, which source redshift is $z_s = 1.034$ (S23).
The shaded regions represent the standard deviation from 10,000 $\kappa$TNG realisations.
All maps are smoothed with a $2 \, \mathrm{arcmin}$ Gaussian window for the PDF, peaks, and minima.}
\label{fig:comparison_all}
\end{figure*}

We compare our results to
previous works, including analytic models for the power spectrum from the HMcode~\citep{Mead2015} and cosmological hydrodynamic simulations,
BAHAMAS~\citep{McCarthy2017,McCarthy2018} and Horizon-AGN~\citep{Dubois2014,Gouin2019}.
We show the comparison for the power spectrum, PDF, peaks, and minima in Figure~\ref{fig:comparison_all}.
All maps are smoothed with a $2 \, \mathrm{arcmin}$ Gaussian window for the PDF, peaks, and minima.
All our comparisons are done for $z_s=1$, which is close to the expected mean redshift of source galaxies expected from Stage-IV surveys~\citep{Euclid2011,LSST2009}.

We compare our results to the \textit{halofit}-based HMcode model using several sets of parameters,
all calibrated against different runs of the OWLS simulations \citep{Schaye2010,vanDaalen2011}, including
(1) DMONLY, in which only gravity from dark matter is implemented and no baryonic physics is included,
(2) REF, in which radiative cooling and heating, star formation and evolution, chemical enrichment,
and supernovae feedback are included,
(3) DBLIM, which adopts a top-heavy initial mass function and stronger supernovae feedback in addition to the REF model, and
(4) AGN, which includes feedback due to active galactic nuclei to the REF model.
In addition, we compare our results to two other hydrodynamic simulations.
For BAHAMAS, there are 10,000 maps at source redshift $z_s = 1$,
where each covers $5 \times 5 \, \mathrm{deg}^2$ sky and
has a resolution of $0.17 \, \mathrm{arcmin}$ per pixel.
In addition to the fiducial model, there are two additional runs, where AGN feedback is
tuned to be more (``high AGN'') or less (``low AGN'') effective.
These simulations are useful to address the impact of AGN feedback.
For Horizon-AGN, there is only one map at $z_s = 1$ and only power spectrum information is available.
Because the simulations adopt different cosmological parameters, it is infeasible to directly compare the statistics.
Therefore we focus only on the ratio of statistics between the hydrodynamic and corresponding DMO simulations.

For the power spectrum, the suppression at scales $1000 \lesssim \ell \lesssim 5000$
seen in $\kappa$TNG is comparable to that from Horizon-AGN, BAHAMAS ``low AGN'', and HMcode REF.
However, at smaller scales, the amplitude of the suppression is larger for $\kappa$TNG,
which is caused by the strong AGN feedback implemented in IllustrisTNG.
This feature is in general agreement with past results of matter power spectrum \citep{Chisari2019}
and baryon fractions in massive halos \citep{vanDaalen2020}.
For the PDF, the overall trends are similar between TNG and BAHAMAS.
The slightly different zero-crossing is likely due to their different cosmologies and feedback models and hence the overall skewness and width of the PDFs.
For peaks and minima, it is interesting to see that the overall baryonic effects are quite similar between the TNG and BAHAMAS,
despite the very different subgrid models implemented in them. In particular, the TNG results are comparable to the ``low AGN'' run.
In addition, \citet{Yang2013} studied the baryonic effects on peak counts by manually boosting the halo concentration by $50\%$.
They found that low peaks are less affected, as these peaks are typically associated with several small haloes along the line-of-sight direction
and sensitive to only the outer regions of these haloes,
while high peaks are sensitive to the inner regions of single massive haloes.
In contrast, we find that low peaks are equally impacted by baryons,
an evidence that baryons are impacting small haloes beyond their virial radii.

Within the current observational limit, we are not able to distinguish the best model among these curves.
However, future data from optical, X-ray, and thermal and kinetic Sunyaev--Zel'dovich observations are expected
to put significantly tighter constraints on the level of baryonic feedback \citep{Battaglia2012,Hill2016,Amodeo2020},
and these constraints can then be compared with the WL data.

\section{Conclusions}
\label{sec:conclusions}
The uncertainty in modelling baryons is already limiting
the cosmological analysis with current generation weak lensing surveys~\citep{DES2018cosmology}.
If left unaccounted for, the effects of baryons will significantly bias
our constraints on dark energy, dark matter, and neutrino mass from upcoming surveys.
In this work, we generate a suite of mock WL maps, the $\kappa$TNG,
by ray-tracing through the IllustrisTNG hydrodynamic simulations.
We produced 10,000 pseudo-independent maps at a wide source redshift range $0.034 \leq z_s \leq 2.568$,
well covering the range probed by existing and upcoming WL surveys.
Furthermore, we also generate the $\kappa$TNG-Dark maps from the corresponding DMO simulations.
We release our mock convergence maps at our website (\url{http://columbialensing.org}).
By comparing the pair of maps with and without baryonic physics,
one can isolate the baryonic effects on the WL statistics.

We find that baryonic processes suppress the WL angular power spectrum by up to $20\%$.
Towards low redshift, the level of suppression is larger,
potentially due to stronger stellar and AGN feedback;
in addition, the onset of the suppression occurs at larger scales, as the same physical object extends over a larger angular scale.
We also observe an enhancement on smaller scales due to radiative cooling.
The overall suppression from $\kappa$TNG is moderate compared to results from other analytic models and simulations,
consistent with previous works comparing their matter power spectra~\citep{Springel2018}.

We also show the effects of baryons on higher-order  weak lensing statistics: the PDF, peak counts, and counts of minima.
These statistics contain rich non-Gaussian information beyond the power spectrum
and have the potential to place a tighter constraint on cosmological parameters with upcoming surveys.
Our work is the first to show detailed redshift-evolution for these statistics.
In general, baryonic processes suppress both the positive and negative tails of all three statistics.
The change is asymmetric on the two tails and hence is not captured by the power spectrum.
In addition, we also find changes in the total number of peaks and minima.
Compared to past works on baryonic effects with weak lensing non-Gaussian statistics,
mainly from the BAHAMAS simulations, the overall trends are surprisingly similar,
despite the very different subgrid models implemented in these simulations.

While contributing to a large body of semi-analytic tools and hydrodynamic simulations
that have already paved the road to study the effect of baryons on weak lensing observables,
our newly developed $\kappa$TNG suite of mock WL maps provide a venue to extend the studies to those requiring mass maps,
such as non-Gaussian weak lensing statistics and machine learning
with convolutional neural networks
\citep{Schmelzle2017,Gupta2018,Ribli2019a,Merten2019,Ribli2019b,Fluri2019}.
Our maps will also be useful for testing pioneering works on analytic models
for higher-order statistics \citep{fan2010,Thiele2020},
which, in the future, could be extended to incorporate baryonic effects.
With the $\kappa$TNG and $\kappa$TNG-Dark map pairs,
one can also learn the DMO-to-hydrodynamic mapping
using the generative adversarial network or variational auto-encorder \citep[see, e.g.,][]{Troster2019}
and augment existing DMO simulations.
It is also possible to explore physically motivated mapping methods,
similar the gradient-descent method adopted by \citet{Dai2018} for the three-dimensional density field.

\section*{Acknowledgements}
We are grateful to Shy Genel for his help with the IllustrisTNG simulations.
We thank Ian McCarthy for sharing the BAHAMAS convergence maps.
We thank Will Coulton, Fran\c{c}ois Lanusse, David Spergel, and Masahiro Takada for useful discussions.
We thank Ryuichi Takahashi for useful comments on the earlier version of the manuscript.
KO is supported by JSPS Overseas Research Fellowships.
This work is in part supported by an NSF Astronomy and Astrophysics Postdoctoral Fellowship (to JL)
under award AST-1602663, and by NASA through ATP grant 80NSSC18K1093 (to ZH).
This work used the Extreme Science and Engineering Discovery Environment (XSEDE),
which is supported by NSF Grant No.~ACI-1053575.

\section*{Data Availability}
The data underlying this article will be partially available at
Columbia Lensing (\url{http://columbialensing.org}), with 100 realisations for three source redshifts ($z_s = 0.5$, $1.0$, $1.5$).
The remaining data will be shared on reasonable request to the corresponding author.
The simulation data of IllustrisTNG is available at \url{https://www.tng-project.org/}.



\bibliographystyle{mnras}
\bibliography{main}



\appendix

\section{Fitting Formula for the Power Spectrum Ratio}
\label{sec:fitting}
\begin{figure}
\includegraphics[width=\columnwidth]{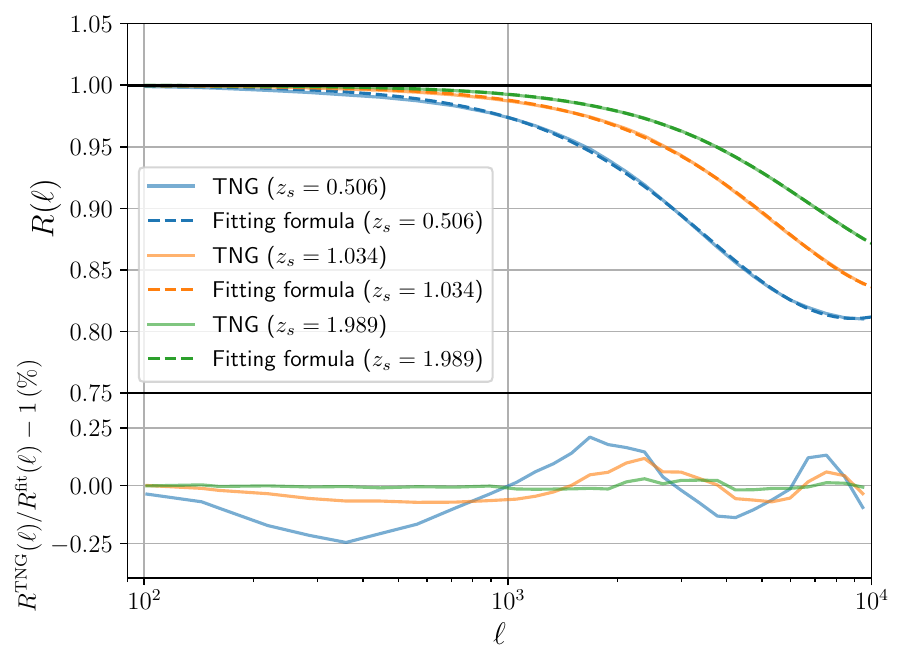}
\caption{\textbf{Upper panel}: Comparison of
angular power spectrum ratios measured from $\kappa$TNG and the fitting formula
for three source redshifts $z_s = 0.506, 1.034, 1.989$.
\textbf{Lower panel}: Fractional differences between $\kappa$TNG and the fitting formula.}
\label{fig:ratio_fitting}
\end{figure}

We provide a fitting formula that allows a quick exploration of baryonic effects on the angular power spectrum:
\begin{align}
  \label{eq:fitting}
  C_\kappa^\mathrm{baryon} (\ell) / C_\kappa^\mathrm{DMO} (\ell) =
  R^\mathrm{fit} (\ell) \equiv \frac{1+(\ell/\ell_{s1})^{\alpha_1}}{1+(\ell/\ell_{s2})^{\alpha_2}} ,
\end{align}
where $\ell_{s1}$, $\ell_{s2}$, $\alpha_1$, and $\alpha_2$ are redshift-dependent free parameters.
Their values are tabulated in Table~\ref{tab:fitting_parameters}.
In optimising the fitting formula, we excluded measurements of $\ell = 3771.0 \text{--} 9465.6$
for the first source redshift $z_s = 0.034$ (S1) because of strong artefacts due to resolution.
In Figure~\ref{fig:ratio_fitting},
we show the comparison of angular power spectrum ratio with $\kappa$TNG
and the fitting formula for $z_s = 0.506$ (S13), $1.034$ (S23), and $1.989$ (S35).
The accuracy of the fitting formula is better than $0.3\%$
for most source redshifts of $\kappa$TNG maps,
but slightly worse for low source redshifts.

\begin{table}
  \caption{The parameters of the power spectrum fitting formula Eq.~\ref{eq:fitting}.}
  \label{tab:fitting_parameters}
\begin{tabular}{cccccc}
Source Plane & $z_s$ & $\ell_{s1}$ & $\ell_{s2}$ & $\alpha_1$ & $\alpha_2$ \\
\hline \hline
S1 & $0.034$ & $582.64$ & $441.63$ & $1.693$ & $1.478$ \\
S2 & $0.070$ & $3773.54$ & $2529.39$ & $1.284$ & $0.939$ \\
S3 & $0.105$ & $2921.87$ & $2104.38$ & $1.459$ & $1.224$ \\
S4 & $0.142$ & $3250.72$ & $2438.22$ & $1.572$ & $1.364$ \\
S5 & $0.179$ & $3724.26$ & $2855.09$ & $1.645$ & $1.452$ \\
S6 & $0.216$ & $4201.42$ & $3267.03$ & $1.710$ & $1.523$ \\
S7 & $0.255$ & $4664.31$ & $3663.60$ & $1.762$ & $1.578$ \\
S8 & $0.294$ & $5119.63$ & $4051.22$ & $1.808$ & $1.624$ \\
S9 & $0.335$ & $5600.60$ & $4456.01$ & $1.846$ & $1.657$ \\
S10 & $0.376$ & $6069.34$ & $4852.78$ & $1.877$ & $1.684$ \\
S11 & $0.418$ & $6559.00$ & $5264.39$ & $1.902$ & $1.703$ \\
S12 & $0.462$ & $7099.52$ & $5712.55$ & $1.923$ & $1.714$ \\
S13 & $0.506$ & $7698.22$ & $6203.65$ & $1.940$ & $1.720$ \\
S14 & $0.552$ & $8318.01$ & $6711.92$ & $1.956$ & $1.722$ \\
S15 & $0.599$ & $8945.57$ & $7227.90$ & $1.970$ & $1.722$ \\
S16 & $0.648$ & $9577.85$ & $7749.58$ & $1.983$ & $1.721$ \\
S17 & $0.698$ & $10217.52$ & $8279.71$ & $1.994$ & $1.718$ \\
S18 & $0.749$ & $10859.85$ & $8815.16$ & $2.005$ & $1.714$ \\
S19 & $0.803$ & $11488.44$ & $9343.56$ & $2.014$ & $1.710$ \\
S20 & $0.858$ & $12086.34$ & $9851.46$ & $2.021$ & $1.706$ \\
S21 & $0.914$ & $12635.92$ & $10323.82$ & $2.026$ & $1.702$ \\
S22 & $0.973$ & $13128.47$ & $10751.96$ & $2.028$ & $1.700$ \\
S23 & $1.034$ & $13551.94$ & $11124.61$ & $2.027$ & $1.698$ \\
S24 & $1.097$ & $13890.45$ & $11427.82$ & $2.023$ & $1.698$ \\
S25 & $1.163$ & $14127.45$ & $11647.41$ & $2.016$ & $1.700$ \\
S26 & $1.231$ & $14257.10$ & $11778.76$ & $2.006$ & $1.704$ \\
S27 & $1.302$ & $14279.51$ & $11822.83$ & $1.993$ & $1.709$ \\
S28 & $1.375$ & $14210.38$ & $11794.04$ & $1.978$ & $1.715$ \\
S29 & $1.452$ & $14075.06$ & $11714.29$ & $1.964$ & $1.721$ \\
S30 & $1.532$ & $13900.02$ & $11605.36$ & $1.950$ & $1.727$ \\
S31 & $1.615$ & $13698.63$ & $11477.60$ & $1.937$ & $1.734$ \\
S32 & $1.703$ & $13478.39$ & $11336.30$ & $1.926$ & $1.740$ \\
S33 & $1.794$ & $13238.19$ & $11179.64$ & $1.915$ & $1.746$ \\
S34 & $1.889$ & $12987.06$ & $11014.19$ & $1.906$ & $1.752$ \\
S35 & $1.989$ & $12729.72$ & $10843.22$ & $1.898$ & $1.758$ \\
S36 & $2.094$ & $12465.76$ & $10665.78$ & $1.890$ & $1.764$ \\
S37 & $2.203$ & $12195.18$ & $10481.52$ & $1.883$ & $1.769$ \\
S38 & $2.319$ & $11919.66$ & $10291.42$ & $1.877$ & $1.774$ \\
S39 & $2.440$ & $11641.09$ & $10096.77$ & $1.871$ & $1.779$ \\
S40 & $2.568$ & $11362.87$ & $9900.14$ & $1.866$ & $1.784$ \\
\hline
\end{tabular}
\end{table}

\bsp 
\label{lastpage}
\end{document}